\documentclass[a4paper,article,twocolumn,english,twocolumns,epl]{revtex4}
\usepackage{amsmath}
\usepackage{amsfonts}
\usepackage{amssymb}
\usepackage{comment}
\usepackage{color}
\usepackage{graphicx}
\providecommand{\U}[1]{\protect\rule{.1in}{.1in}}
\begin{document}
\title{Fluctuation Induced Forces in Non-equilibrium (Diffusive) Dynamics}
\date{\today}
\author{Avi Aminov}
\affiliation{Department of Physics, Technion, Haifa 32000, Israel}
\author{Mehran Kardar}
\affiliation{Department of Physics, Massachusetts Institute of
Technology, Cambridge, Massachusetts 02139, USA}
\author{Yariv Kafri}
\affiliation{Department of Physics, Technion, Haifa 32000, Israel}

\begin{abstract}
Thermal fluctuations in  non-equilibrium steady states generically lead to power law decay of correlations for conserved quantities.  
Embedded bodies which constrain fluctuations in turn experience fluctuation induced forces. 
We compute these forces for the simple case of parallel slabs in a driven diffusive system.
The force falls off with slab separation $d$ as $k_BT/d$ (at temperature $T$, and
in all spatial dimensions),
but can be attractive or repulsive. 
Unlike the equilibrium Casimir force, the force amplitude is non-universal and explicitly depends on dynamics.
The techniques introduced can be generalized to study pressure and fluctuation induced forces in a broad class of non-equilibrium systems.
\end{abstract}

\pacs{05.40.-a, 05.10.Gg}

\maketitle

External objects immersed in a medium typically modify the underlying fluctuations
and in turn experience fluctuation--induced force (FIF)~\cite{Golestanian99}.
The textbook example is the Casimir force~\cite{Casimir,rev1} arising from quantum
fluctuations of the electromagnetic field;
its thermal analog in critical systems~\cite{FG} has also been observed 
in binary liquid mixtures~\cite{Bechinger}.
In both cases, the underlying fluctuations are long-range correlated leading to forces
that fall off as power laws.
In the latter (oil/water mixture) this is achieved by tuning to a critical point,
while the former is a consequence of the massless nature of the photon field.
Generically in a fluid in equilibrium, correlations (and hence FIF) decay exponentially
and are insignificant beyond a correlation length.

Non-equilibrium situations provide another route to long-range correlated fluctuations:
Systems which in equilibrium have zero or short-ranged correlations 
($C_{eq}\sim \delta^s({\bf x})$ in $s$  dimensions), quite generically exhibit power law
correlations ($C_{neq}\sim 1/|{\bf x}|^s$) with conserved dynamics when out of equilibrium~\cite{GLS}.
It is thus natural to inquire about the nature (strength and range) of FIF in corresponding
non-equilibrium settings (where there is no corresponding force in equilibrium).
Such forces have indeed been explored in a number of circumstances, including
driven granular fluids~\cite{Cattuto06,Cattuto07,wolf}, shear flow \cite{Wada03}, and in ordinary fluids
subject to a temperature gradient~\cite{Sengers13,Sengers14}.
Here, we explore possibly the simplest (and hence analytically tractable) 
example of FIF in a system of diffusing particles which are subject to hard core exclusion, commonly referred to as the symmetric simple exclusion process (SSEP)~\cite{Derrida07}. 

\begin{figure}
\includegraphics[width=0.5\textwidth]{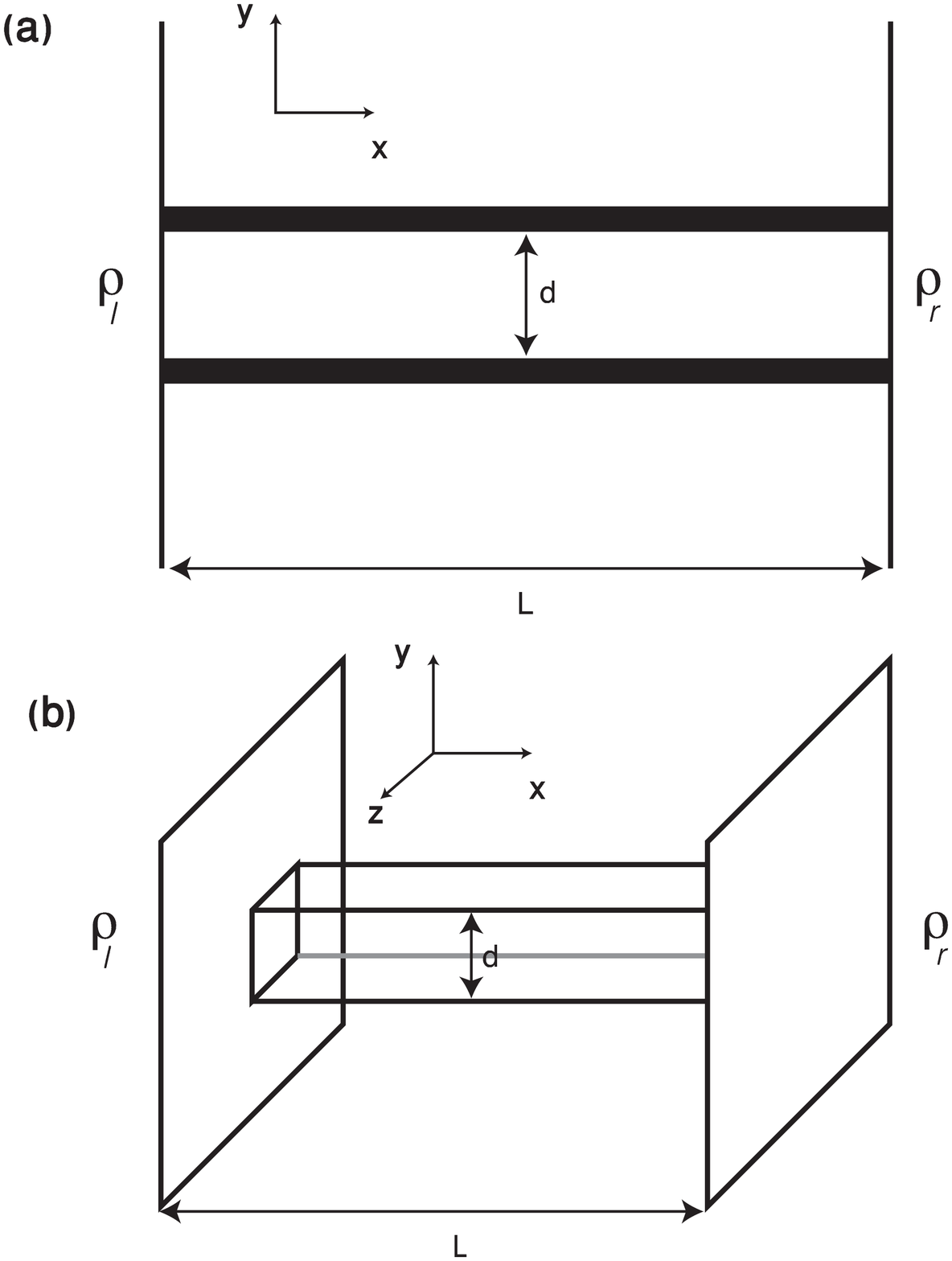}
\begin{center}
\caption{The setups studied consist of:
{\bf (a)} A two dimensional system, infinite in the $y$ direction is connected to two reservoirs at $x=0$ and $x=L$,
with densities   $\rho(0,y)=\rho_l$ and $\rho(L,y)=\rho_r$, respectively.
Two slabs, a distance $d$ from each other, span the system along the $x$ direction. 
{\bf (b)} The three dimensional generalization of the above, 
with the two slabs  replaced by a tube of square cross section. \label{fig:2dsetup}} \label{setup}
\end{center}
\end{figure}

The setups examined are: {\bf (a)} The two dimensional system shown in Fig.~\ref{fig:2dsetup}(a); infinite in the $y$ direction and connected to two reservoirs at $x=0$ and $x=L$,  with densities  $\rho(0,y)=\rho_l$ and $\rho(L,y)=\rho_r$, respectively.
Two slabs, a distance $d$ from each other, span the system along the $x$ direction. 
{\bf (b)} The three dimensional extension of this setup depicted in Fig.~\ref{fig:2dsetup}(b), with the two slabs  replaced by a tube of square cross section. 
{\bf (c)} A generalized setup in which the slabs (or tube in three dimensions) of  length $R\leq L$,  do not 
necessarily span the entire system.

Consider first the two dimensional setup of Fig.~\ref{fig:2dsetup}(a). For equal reservoir densities, $\rho_l=\rho_r$, the system is in equilibrium, the pressure is uniform throughout the box and there is no average force on the slabs. When the reservoir densities are different, 
the (average) density profile varies linearly between the two reservoirs,
and there is an average diffusive current of particles along the $x$ direction. 
Its magnitude is  $j=D\Delta\rho/L$, where $D$ the diffusion constant of the particles and $\Delta\rho\equiv \left(\rho_l-\rho_r\right)$. Since the average density profile is the same on both sides of each slab,
naively one would again expect no force between the two plates. 
However, we find that the presence of non-equilibrium long--range correlations~\cite{Derrida07,Spohn83}
for $\rho_l \neq \rho_r$   leads to a force between the two slabs, given by (for $d \ll L$)
 \begin{eqnarray}
F &=& -\frac{k_BT}{d} (\Delta\rho)^2 g(\rho_l,\rho_r) \nonumber \\  
 &=&  -\frac{k_BT}{d}\left(\frac{jL}{D}\right)^2 g(\rho_l,\rho_r) \label{eq:Force} \;.
\end{eqnarray}
Here $k_B$ is the Boltzmann constant, $T$ is the temperature of the surrounding bath and $g(\rho_l,\rho_r)$ is a positive dimensionless function of order one. Note that the force is attractive and when expressed in terms of current, or the average density gradient $\overline{ \nabla \rho }= \Delta\rho/L$, proportional to $L^2 \left(\, \overline{\nabla \rho}\, \right)^2 $. Here the overline denotes an average over the steady--state probability distribution.  When the three dimensional analogue of the above setup is considered and a tube with a square cross section connects the two reservoirs (see Fig.~\ref{setup}(b)) the force between two parallel slabs has the same form, with the same function $g(\rho_l,\rho_r)$. For $d \gg L$ the force still decays as $1/d$ but with a coefficient that is smaller by a factor of 2.

While the force is attractive for SSEP, it can be repulsive in other interacting diffusive systems. 
Using a simplified model, we argue that this is the case in boundary driven antiferromagnetic Ising models with spin conserving  dynamics for a certain regime of parameters. 

Finally, our results suggest that when the slabs or tube is of finite extension, $R$, the force should behave as 
\begin{eqnarray}
F &=& -\frac{k_BT}{d} \frac{R^2}{L^2}(\Delta\rho)^2 g^{\star}\left(\rho_l,\rho_r,\frac{R}{x_0}\right) \nonumber \\ 
&=&-\frac{k_BT}{d}R^2 \left(\frac{j}{D}\right)^2  g^{\star}\left(\rho_l,\rho_r,\frac{R}{x_0}\right) \;.
\label{eq:forceR}
\end{eqnarray}
Here $g^{\star}$ is a positive function of $\rho_l$, $\rho_r$ and $R$, while $x_0$ is the distance of the slabs from the left reservoir. 
For hard core particles the force is attractive and proportional to $R^2$.
 A similar scaling form, but with  opposite sign, is expected for the boundary driven antiferromagnetic Ising model.

To derive the above results we use the formalism of fluctuating hydrodynamics~\cite{Derrida07,Spohn83,SengersBook06}.  In this approach the dynamical equation of motion for the particle density can be shown, either through a microscopic derivation (for example, see~\cite{Tailleur08}) or through a phenomenological approach,  to be
\begin{equation}
\partial_{t}\rho\left(x,t\right)+\partial_{x}J\left(x,t\right)=0\;,\label{eq:continuity}
\end{equation}
with a stochastic current
\begin{equation}
J_{\mu}\left(x,t\right)=-D\partial_\mu \rho\left({\bf x},t\right)+\sqrt{\sigma\left(\rho\right)}\eta_{\mu}\left({\bf x},t\right)\;.\label{eq:current_dynamics}
\end{equation}
Here, $D$ is a diffusion coefficient and $\eta_{\mu}$ is an uncorrelated white noise vector with components $\mu=1, \cdots, s$, where $s$ the system dimension. The noise has zero mean $\overline{ \eta_\mu \left({\bf x},t\right)} =0$, is uncorrelated $\overline{ \eta_\mu \left({\bf x},t\right)\eta_\nu \left({\bf x'},t'\right)} =\delta_{\mu,\nu}\delta\left(t-t'\right)\delta\left({\bf x}-{\bf x}'\right)$; its variance $\sigma(\rho)=2Dk_{B}T\rho^{2}\kappa(\rho)$  satisfying a fluctuation--dissipation condition, where $\kappa(\rho)$ is the compressibility of the gas. For diffusing particles subject to hard-core exclusion, $D$ is a constant independent of the density $\rho$, and $\sigma(\rho)=2D a^s \rho(1-\rho)$~\cite{Derrida07,Spohn83}. Here $a$ is a UV cutoff given by the lattice size and we use the standard convention where  $0 \leq \rho \leq 1$ is dimensionless.
For simplicity, in what follows derivations are mostly restricted to the two dimensions (Fig.~\ref{setup} (a));
the extension to  three dimensions is straightforward and for we only quote the final results. 

The density is subject to the boundary conditions $\rho (0,y)=\rho_l$ and $\rho (L,y)=\rho_r$ at the reservoris,
while the normal component of the current must vanish on the two slabs.
In steady-state  the average density density profile is given by $\overline{\rho} (x,y) = \rho_l+\Delta\rho\,x/L$,
with $\overline{{\bf j}}=(D\Delta\rho/L)\hat{\bf x}$.

It is important to note that the continuum equations are valid in the hydrodynamic limit 
of a corresponding lattice obtained as follows: 
Consider a (hyper-)cubic system of volume $L^s$ 
divided into $N^s$ boxes of size $\xi^s$, where $\xi$ is a length scale such that $N\xi=L$. 
The hydrodynamic regime corresponds to first letting $\xi \to \infty$ with $L/ \xi= N$, and then taking the limit $N \to \infty$. Equation~\ref{eq:continuity} is valid when the system is rescaled and length is measured in units where $\xi \to 0$ and $N\xi=L$~\cite{Derrida07}. 

With this in mind and using ideas similar to Refs.~\cite{Cattuto06,Cattuto07,Sengers13,Sengers14} we write the average pressure to leading order in the fluctuations as:
\begin{equation}
\overline{ P(\rho({\bf x})) }=\lim_{\xi \to \infty} \left(P( \overline{\rho}({\bf x}) )+ \frac{1}{2} \left.P''\right|_{\overline{\rho}({\bf x})}\overline{\delta\rho{\left({\bf x}\right)}^2} \right)\;.
\label{eq:pressure}
\end{equation}  
Here, $\delta \rho({\bf x}) = \rho({\bf x}) - \overline{\rho}({\bf x})$, and primes henceforth indicate derivatives with respect to the density $\rho$. To calculate the force between the plates the pressure has to be evaluated on both sides of each slab. The hydrodynamic procedure described above implies that calculations have to be carried out using Eq.~\ref{eq:continuity}  with the cutoff $\xi$, and then with length scales rescaled at the end of the calculation so that $L$ is finite.  In practice this implies that any divergent UV contributions to the pressure fluctuations need to be removed from the results of the calculation. In particular, in equilibrium and using the continuum result  $\overline{ \delta \rho({\bf x}) \delta \rho({\bf x}') }=a^s\rho(1-\rho) \delta({\bf x}-{\bf x}')$, one has $\overline{ \delta \rho({\bf x}) ^2 }= a^s\rho(1-\rho)/\xi^s$,  and the fluctuations do not contribute to the pressure as $\xi\to\infty$.

Clearly, in the setup considered, at any location along the wall contributions from $P(\overline{\rho}({\bf x}))$ cancel. However, as we now show $\overline{\delta \rho({\bf x}) ^2}$ varies on the opposing faces of each slab leading to a fluctuation induced force. 
To evaluate the out of equilibrium fluctuation induced contribution to pressure, note that for the SSEP the fluctuation--dissipation relation, with $\kappa(\rho)=\frac{1}{\rho}\frac{ d \rho}{d P}$ gives
\begin{equation}
\frac{1}{2}\left.P''\right|_{\overline{\rho}({\bf x})}= \frac{1}{2a^s} \frac{k_BT}{(1-\overline{\rho}({\bf x}))^2} \;.
\label{eq:p"}
\end{equation}
To compute $\overline{ \delta \rho({\bf x}) ^2 }$ on the faces of the slabs, we evaluate the fluctuations along the walls in chambers of size $L \times d$ and  $L \times \infty$, respectively. The first corresponds to the chamber  between the walls,
and the second to the semi-infinite surrounding spaces.

To evaluate $\overline{\delta \rho(x,y) ^2 }$ we use standard methods~\cite{Gardiner94,Garcia87}, expanding the equation of motion to linear order in $\delta \rho$ about the steady-state profile. To linear order the current is
\begin{equation}
J_{\mu}\left(x,t\right)=-D\partial_{\mu} \delta \rho\left({\bf x},t\right)+\sqrt{\sigma\left(\overline{\rho}({\bf x}) \right)}\eta_{\mu}\left({\bf x},t\right).\label{eq:current_dynamics_delta_rho}
\end{equation}
The dynamical equation is then linear in $\delta \rho$ so that the correlation function $C({\bf x},{\bf x}')=\overline{ \delta \rho({\bf x})\delta \rho({\bf x}') }$ satisfies a Lyapunov equation~\cite{Gardiner94,Garcia87}. After several straightforward manipulations this can be brought to the form
\begin{eqnarray}
(\nabla_{\bf x} D  \nabla_{\bf x}&+&\nabla_{{\bf x}'} D \nabla_{{\bf x}'})C_{neq}({\bf x},{\bf x}')\nonumber \\ &=&-\frac{1}{2}\delta({\bf x}-{\bf x}') \nabla^2_{{\bf x}'}\sigma(\overline{\rho}({\bf x}') )\,,
\label{eq:Lyapunov}
\end{eqnarray}
where $C_{neq}({\bf x},{\bf x}')=C({\bf x},{\bf x}')-\frac{1}{2D}\sigma(\overline{\rho}({\bf x}')) \delta({\bf x}-{\bf x}')$ is the non-equilibrium part of the correlation function. Using  the average density profile, $\overline{\rho} (x,y) = \rho_l+\Delta\rho\,x/L$, the above equation reduces to calculating the Green's function of a Poisson equation:
\begin{equation}
(\nabla^2_{\bf x}+\nabla^2_{{\bf x}'})C_{neq}({\bf x},{\bf x}')= 2\delta({\bf x}-{\bf x}') (\Delta\rho)^2a^2/L^2\;.
\label{eq:Poisson}
\end{equation}
The boundary conditions are such that $C_{neq}=0$  when either ${\bf x}$ and ${\bf x}'$ are on the reservoirs 
(since the density  on the reservoirs is fixed, $\delta \rho=0$ identically), while on the slabs its normal
derivative is zero (no current).  
To calculate the force, density fluctuations have to be calculated on the slabs, e.g. 
$c_{neq}(x)\equiv C_{neq}(\lbrace x,y=0\rbrace,\lbrace x, y=0\rbrace)$, evaluated at the same point 
${\bf x}={\bf x}'$  on one of the slabs.
Using standard Fourier methods one finds
\begin{equation}
c_{neq}(x)=\sum_{n} A_n \sin^2\left(\frac{n\pi}{L}x\right) \,,
\label{full_summation}
\end{equation}
with
\begin{equation}
A_n=-\frac{a^2(\Delta\rho)^2}{Ld} \left[ \left(\frac{1}{n \pi}\right)^2 \!\!+\frac{d}{n\pi L}\coth \left( \frac{n\pi  d}{L}\right)   \right] \,. \label{2d}
\end{equation}
In the limit $d \gg L$, 
one  finds to  order  $L/d$
\begin{equation}
A_n=-a^2(\Delta\rho)^2\left[\frac{1}{( n\pi) L^2}+\frac{1}{(n\pi )^2Ld}\right] \;.
\label{eq:dLarge}
\end{equation}
Conversely, for $d \ll L$ (indicated by the superscript $1$) and to leading order in $d/L$
\begin{equation}
A^1_n=-a^2\frac{2(\Delta\rho)^2}{Ld}  \left(\frac{1}{n \pi}\right)^2 \;.
\label{eq:dSmall}
\end{equation}

The Fourier series with $A_n\propto (n\pi)^{-2}$ corresponds to a parabola. For $d \ll L$ this gives
\begin{equation}
c_{neq}^1(x)=-a^2\frac{(\Delta\rho)^2}{Ld}  \frac{x}{L} \left( 1-\frac{x}{L}\right)\;,
\label{eq:autocorr-approximation}
\end{equation}
which is in fact the expected behavior of a one-dimensional SSEP~\cite{Derrida07,Spohn83}.
For $d\gg L$, Eq.~\ref{eq:dLarge} leads to a constant contribution, corresponding to the $d \to \infty$ limit, and a contribution similar to $c_{neq}^1(x)$  with a co-efficient
that is smaller by 2. Using the hydrodynamic procedure described 
earlier, we observe that $\overline{\delta \rho ( \lbrace{ x,y=0 \rbrace})^2} =c_{neq}(x)$. Namely, only the long-range part of the correlation function contributes to the pressure.

The fluctuation--induced correction to the pressure in Eq.~\ref{eq:pressure} is
the product of two factors: the first (given in Eq.~\ref{eq:p"}) is positive,
while the second (from Eq.~\ref{eq:autocorr-approximation}) is negative.
This leads to a negative contribution to pressure, corresponding to attraction between the slabs. In the limit $d \ll L$ the contribution from the semi-infinite surrounding spaces is negligible.
Integrating the local pressure over the slab leads to a fluctuation-induced force 
\begin{eqnarray}
F &=& \int{\mathrm d}x\, \frac{1}{2}\left.P''\right|_{\overline{\rho}({\bf x})} c_{neq}^1(x) \label{eq:force2da} \\  
 &=& -\frac{k_BT(\Delta\rho)^2}{d}\int_0^1 dz  \frac{z\left( 1-z \right)}{2(1- \overline{\rho}(z) )^2}  \;,
\label{eq:force2d}
\end{eqnarray}
as proposed in Eq.~\ref{eq:Force}. Here $\overline{\rho}(z) =\rho_l+\Delta\rho\,z$. Evaluating the integral shows that the total force is a concave function,
vanishing at $\rho_l=\rho_r$. It is straightforward to use the above results to verify that in the limit $d \gg L$ the force decays in the same form with a co-efficient that is smaller by 2.
The calculation can be repeated in three dimensions for the configuration depicted in Fig.~\ref{fig:2dsetup}. The force is now calculated between two opposite slabs, say in the $y$ direction and yields the exact same result as above.

\begin{figure}
\begin{center}
\includegraphics[width=0.5\textwidth]{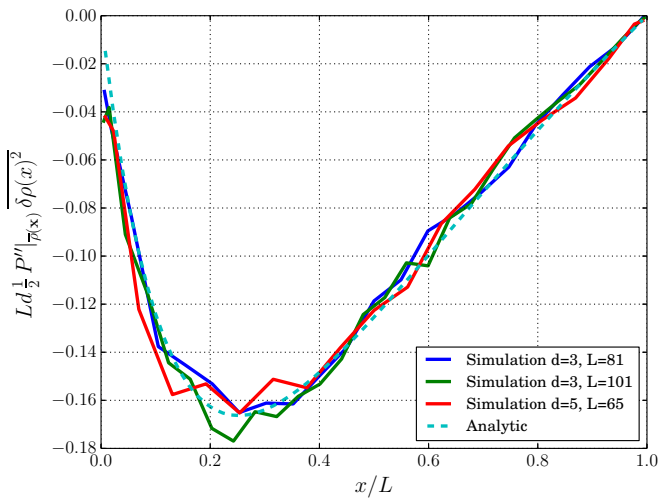}
\caption{\label{fig:numerics} Numerical results for the fluctuation induced pressure in two-dimensions, given by the integrand of Eq.~\ref{eq:force2da}, multiplied by $Ld$.  Here $\rho_L=0.1$, $\Delta\rho=0.6$ and three values of $L,\,d$ such that $d \ll L$ are shown. The units are chosen such that the lattice spacing is set to  $a=1$ and $k_BT=1$. The solid lines depict  numerical results, while the dashed line is the analytic calculation. The numerical method  for measuring the pressure is described in Appendix~\ref{sec:appendix-numerical-method}.}
\end{center}
\end{figure}

The negative result in Eq.~\ref{eq:autocorr-approximation} may appear counterintuitive, since it originates from a computation of $\overline{\delta\rho{\left(x\right)}^2}$. 
To validate this conclusion, and the underlying 
hydrodynamic procedure, we performed Monte--Carlo simulations on a two-dimensional square lattice and measured the pressure along the slab (see Appendix~\ref{sec:appendix-numerical-method} for details). The results in the limit $d/L \ll 1$ and for different lattice sizes are shown in Fig.~\ref{fig:numerics}. The numerics compare well with the theoretical predictions. 

Equation~\ref{eq:pressure} suggests that the pressure, and therefore the force, can be either positive or negative,
depending on the relative signs of $P''$ and $c_{neq}$.
To explore this further we carry out a perturbation theory in $\Delta \rho$ for a general model with a density dependent diffusion constant $D(\rho)$. The equation for the average density is then  $\nabla \left(D(\overline{\rho}({\bf x})) \cdot \nabla \overline{\rho}({\bf x})\right)=0$, and the Lyapunov Eq. \ref{eq:Lyapunov} now has $D$ as a function of $\rho$. Setting $\overline{\rho}(x)=\rho_l+\rho_1(x)\Delta \rho + \rho_2(x) (\Delta \rho)^2 +\cdots$, it is straightforward to show that to order $(\Delta \rho)^2$ the final result in Eq.~\ref{eq:autocorr-approximation} is replaced by
\begin{equation}
c_{neq}^1({\bf x}) \simeq \frac{k_BT(\Delta\rho)^2}{2Ld} 
\left[\left(\frac{\rho}{P'}\right)''+\left(\frac{\rho}{P'}\frac{D'}{D}\right)'\right] 
\frac{x}{L} \left(1-\frac{x}{L}\right)\;,
\label{eq:autocorr-general}
\end{equation}
resulting in a force
\begin{equation}
F\simeq\frac{k_BT(\Delta \rho)^2}{24d} P''\left[\left(\frac{\rho}{P'}\right)''+\left(\frac{\rho}{P'}\frac{D'}{D}\right)'\right] \;,
\label{eq:force-general}
\end{equation}
where the derivatives with respect to the density are evaluated at $\rho_l$. The second term on the right-hand-side shows the explicit dependence of the results on the dynamics through the appearance of the diffusion coefficient. Moreover,
 there are no apparent restrictions on the sign of the force. Consider for example a model with $D=k^2(1-q^2(\rho-\rho_0))$, $\sigma(\rho)=r^2(1+t^2(\rho-\rho_0)^2)$ and boundary conditions with $\rho_l=\rho_0$. While we are not aware of a direct microscopic realization of this formula, it can be considered as an approximation for an Ising model with repulsive interactions evolving under Kawasaki dynamics, with $\rho$ denoting, say, the density of down spins. There, it is known that in one dimension $\sigma(\rho)$ has a minimum around some $\rho_0$ which depends on the parameters of the model, with $D(\rho)$ peaked around $\rho_0$~\cite{Hager02,Bunin13}.  
(On general grounds this behavior is expected to persist in higher dimensions.) Using the above expressions it is straightforward to check that to order $(\Delta \rho)^2$ the fluctuation induced force, $F\simeq \frac{k_B T (\Delta \rho)^2 t^2}{12 d}$, is repulsive.

The non-extensivity of the force in Eq.~\ref{eq:force2d} is somewhat surprising, and different from say the critical Casimir force
which behaves as $F\propto k_BT L^{s-1}/d^s$ for generalized slabs of side $L$ in $s$ dimensions~\cite{Golestanian99}. 
This is because  $c_{neq}$ scales inversely with the volume of the confining box, resulting in a local pressure
that vanishes for a large slab. As such, we expect this force to be more relevant to small inclusions as opposed to macroscopic slabs.
While the exact solution of the force between two inclusions is beyond the scope of this paper, we can provide an estimate
based on dimensional grounds. 
To this end, we consider parallel slabs of dimension $R$, and neglect the fluctuations of density  at the open sides of the corresponding
enclosure. One is then left with evaluating the pressure fluctuations in a chamber of size $R \times d^{s-1}$ with boundary densities specified by the mean density at the edges of the slab. It is then straightforward to see that in the limit $d \ll L$ the force is now given by (for SSEP)
\begin{equation}
F=-\frac{2k_BT(\Delta\rho)^2}{d}\frac{R^2}{L^2}\int_0^1 dz  \frac{z\left( 1-z \right)}{(1-\overline{\rho}(z))^2}  \;,
\label{eq:force2dR}
\end{equation}
{\it irrespective of dimension $s$},
where $\overline{\rho}(z)=\rho_l+(\Delta\rho)(z_0+Rz)/L$ as advertised in Eq.~\ref{eq:forceR}.

\begin{acknowledgments}
We thank  M. Kolodrubetz and A. Polkovnikov for valuable discussions and suggestions.  
AA and YK are supported by BSF and ISF grants. MK is supported by the NSF through grant No. DMR-12-06323. 
\end{acknowledgments}

\appendix

\section{Numerical evaluation of pressure}
\label{sec:appendix-numerical-method}
To measure pressure in a confined SSEP, we perform Monte-Carlo simulation on a square lattice of size $L\times\left(d+1\right)$, with lattice constant set to one. The lattice sites are then labelled by $(n_x, n_y)$ with $n_x=1,2, \cdots, L$ and  $n_y=1,2, \cdots ,d+1$. The pressure of  hard--core particles is purely entropic, and we measure it 
by a standard method used for evaluating entropic pressure in confined polymers~\cite{dickman87}: To evaluate pressure at site $n'_x$ {\it on a wall}, we introduce a repulsive potential, $V=-k_BT\log\lambda$ on lattice site $(n'_x,d+1)$. 
For the remaining sites on this row, $(n'_x,d+1)$ with $n'_x \neq n_x$, we set $V =\infty$. The pressure is then obtained as
\begin{equation}
P \left(n'_x\right) = \int_0^1 {\mathrm d}\lambda \,\frac{\rho_{d+1}\left(n'_x,\lambda\right)}{\lambda} \;,
\end{equation}
where $\rho_{d+1}(n'_x,\lambda)$ is the average density on site $(n'_x,d+1)$. The integral is performed numerically by discretizing $\lambda\in\left[0,1\right]$ into $20$ equally spaced values. Monte-Carlo simulations are carried out for each value of $n'_x$ and $\lambda$.

The Monte-Carlo simulations are carried out using standard methods: Each lattice site is either occupied  or empty. At each Monte-Carlo step a site $(n_x, n_y)$ is chosen at random. If the site is occupied and $n_x \neq 1,L$, $n_y \neq d,d+1$ an attempted move of the particle is made to one of its randomly chosen nearest--neighbors (with rate 1 in arbitrary units), as long as the hard-core constraint is not violated.
If $n_x=0$ ($n_x=L$), namely near the left (right) reservoir, and the site is empty, a particle is added with rate $\alpha=\frac{\rho_l}{1-\rho_l}$ ($\delta=\frac{\rho_r}{1-\rho_r}$). If the site is occupied, an attempted removal of the particle is made, with equal rate as an attempted move to one of the nearest-neighbors. This choice corresponds to setting $k_BT=1$. Finally, if $n_y=d$ or $d+1$ a move is attempted with rate $\min \lbrace 1,e^{-\beta \Delta V}\rbrace$ where $\Delta V$ is the difference in potential before and after the move. The results presented in the main text (Fig.~\ref{fig:numerics}) were obtained using $8\times 10^{11}$ Monte-Carlo sweeps.

\end{document}